\documentclass[english]{nature}
\usepackage{babel}
\usepackage{amssymb,amsmath,amsfonts}
\usepackage{graphicx}

\usepackage[usenames,dvipsnames]{color}

\makeatletter
\usepackage[mathlines]{lineno}

\def\ket#1{|\mbox{$#1$}\rangle}

\def\bracketii#1#2#3{\langle\mbox{$#1$}|\mbox{$#2$}|\mbox{$#3$}\rangle}

\makeatother

\begin{document}

\title{Quantum nondemolition measurement of an electron spin qubit}

\author{Takashi Nakajima$^{1\dagger*}$, Akito Noiri$^{1\dagger}$, Jun Yoneda$^{1}$, Matthieu R. Delbecq$^{1\S}$, Peter Stano$^{1,2}$, Tomohiro Otsuka$^{1,3\sharp}$, Kenta Takeda$^{1}$, Shinichi Amaha$^{1}$, Giles Allison$^{1}$, Kento Kawasaki$^{4}$, Arne Ludwig$^{5}$, Andreas D. Wieck$^{5}$, Daniel Loss$^{1,6}$ \& Seigo Tarucha$^{1,4*}$}
\maketitle
\begin{affiliations}
\item Center for Emergent Matter Science, RIKEN, 2-1 Hirosawa, Wako-shi, Saitama 351-0198, Japan
\item Institute of Physics, Slovak Academy of Sciences, 845 11 Bratislava, Slovakia
\item JST, PRESTO, 4-1-8 Honcho, Kawaguchi, Saitama, 332-0012, Japan
\item Department of Applied Physics, University of Tokyo, 7-3-1 Hongo, Bunkyo-ku, Tokyo 113-8656, Japan
\item Lehrstuhl f\"{u}r Angewandte Festk\"{o}rperphysik, Ruhr-Universit\"{a}t Bochum, D-44780 Bochum, Germany
\item Department of Physics, University of Basel, Klingelbergstrasse 82, CH-4056 Basel, Switzerland
\item[$\dagger$] These authors contributed equally to this work.
\item[*] Correspondence should be addressed to nakajima.physics@icloud.com or tarucha@ap.t.u-tokyo.ac.jp

\item[$\S$] Present address: Laboratoire Pierre Aigrain, Ecole Normale Sup\'{e}rieure-PSL Research University, CNRS, Universit\'{e} Pierre et Marie Curie-Sorbonne Universit\'{e}s, Universit\'{e} Paris Diderot-Sorbonne Paris Cit\'{e}, 24 rue Lhomond, 75231 Paris Cedex 05, France
\item[$\sharp$] Present address: Research Institute of Electrical Communication, Tohoku University, 2-1-1 Katahira, Aoba-ku, Sendai 980-8577, Japan
\end{affiliations}

\begin{abstract}
Measurement of quantum systems inevitably involves disturbance in various forms.
Within the limits imposed by quantum mechanics, however, one can design an ``ideal'' projective measurement that does not introduce a back action on the measured observable, known as a quantum nondemolition (QND) measurement\cite{Grangier1998,Imoto1985}.
Here we demonstrate an all-electrical QND measurement of a single electron spin in a gate-defined quantum dot via an exchange-coupled ancilla qubit\cite{Mehl2015a,Noiri2017a}.
The ancilla qubit, encoded in the singlet-triplet two-electron subspace, is entangled with the single spin and subsequently read out in a single shot projective measurement at a rate two orders of magnitude faster than the spin relaxation.
The QND nature of the measurement protocol\cite{Lupascu2007,Jiang2009} is evidenced by observing a monotonic increase of the readout fidelity over one hundred repetitive measurements against arbitrary input states. We extract information from the measurement record using the method of optimal inference, which is tolerant to the presence of the relaxation and dephasing.
The QND measurement allows us to observe spontaneous spin flips (quantum jumps)\cite{Vamivakas2010} in an isolated system with small disturbance.
Combined with the high-fidelity control of spin qubits\cite{Veldhorst2014,Kawakami2014,Takeda2016,Yoneda2017,Veldhorst2015,Zajac2017,Watson2017}, these results pave the way for various measurement-based quantum state manipulations including quantum error correction protocols\cite{Nielsen2010,Ralph2006}.
\end{abstract}

Spin-based qubits in semiconductor quantum dots proposed by Loss and DiVincenzo\cite{Loss1998} are a promising platform for universal quantum computing due to high-fidelity control of their coherent states\cite{Veldhorst2014,Kawakami2014,Veldhorst2015,Takeda2016,Yoneda2017,Zajac2017,Watson2017} and the industry-compatible scalable architecture.
One of the current bottlenecks for the single-electron spin qubit is the fidelity and the speed of its initialization and measurement. These limitations are posed by the inherent destructiveness of the currently used single-shot measurement method\cite{Elzerman2004}.
A QND measurement offers unique possibilities to overcome the limitations such as repetitive readout\cite{Jiang2009} and feedback-controlled initialization\cite{Riste2012a}.
The QND measurement has remained elusive for electron spins in contrast to other solid-state systems such as superconducting qubits\cite{Lupascu2007}, or nuclear spins in diamond color centers\cite{Neumann:2010kx,Robledo2011} and in silicon donors\cite{Pla2013}.
While particular types of the photonic readouts of electron spins\cite{Vamivakas2010,Mi2017,Samkharadze2017} can be, in principle, QND, their QND nature has not been demonstrated so far.
Moreover, the QND measurement \emph{via an ancillary qubit} is crucial\cite{Ralph2006} for realizing measurement-based quantum algorithms including quantum error correction codes.

\begin{figure}
\begin{centering}
\includegraphics[width=1\textwidth]{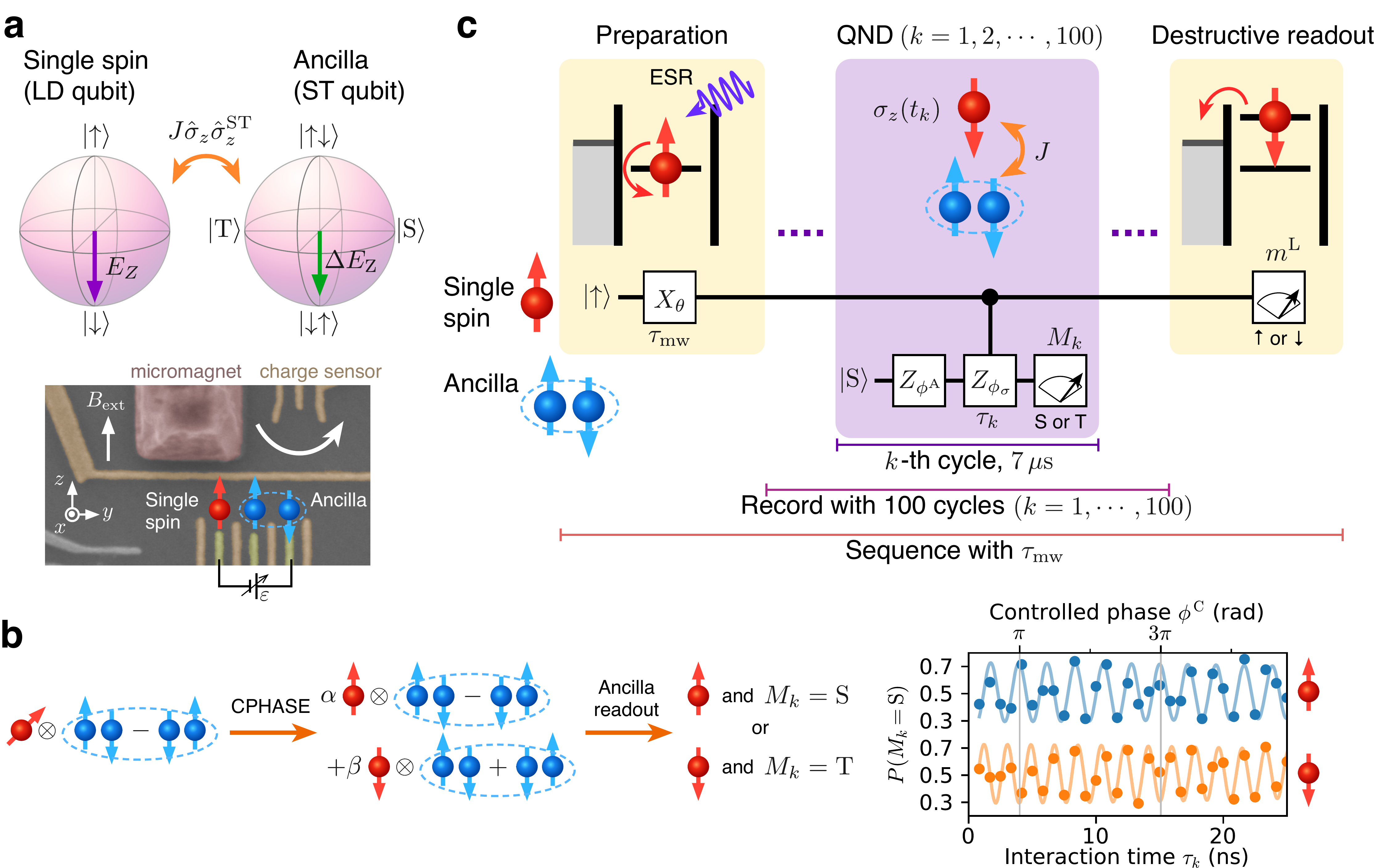}
\par\end{centering}

\protect\caption{\textbf{QND readout of a single electron spin via an ancilla ST qubit.}
\textbf{a,}
Schematic of the triple quantum dot device made of a single electron spin (LD qubit) and an ancilla (ST qubit) coupled by the exchange interaction $J$ (upper panel) and false-colored scanning electron micrograph image of the TQD device (lower panel). The cobalt micromagnet deposited on the wafer surface is magnetized by an in-plane magnetic field $B_\text{ext}=3.155\,\text{T}$ to induce the Zeeman field gradient $\Delta E_\text{Z}$ and the slanting field for the MM-ESR. The TQD charge configuration is probed by a proximal quantum dot charge sensor using radio-frequency reflectometry\cite{Reilly2007,Barthel:2010fk}.
\textbf{b,}
Schematic of the QND measurement protocol (left panel) and singlet-triplet precession of the ancilla showing the spin-dependent phase (right panel). The electron spin (red) is entangled with the ancilla (blue) by a controlled-$Z$ rotation and then the ancilla is projected onto either singlet (S) or triplet (T) state. Here we illustrate a case where the total phase of the ancilla is $\phi_\uparrow + \phi^\text{A} = 0$ for an up-spin state and $\phi_\downarrow + \phi^\text{A} = \pi$ for a down-spin state (the controlled phase is $\phi^\text{C}=\phi_\downarrow-\phi_\uparrow=\pi$). The right panel shows the oscillations of the singlet probability for the ancilla measured with the up- and down-spin input states.
\textbf{c,}
Experimental sequence for repeated QND measurement and a subsequent destructive readout of the electron spin. After preparing an input spin state $\sigma_z(0)$ by the MM-ESR, QND readout cycles with varied interaction time $\tau_k$ are repeated for $100$ times, each of which takes $7\,\mu\text{s}$. Finally, the LD qubit state is read out by energy-selective tunneling\cite{Elzerman2004}.
}
\label{fig:system}
\end{figure}

Here, we demonstrate the QND measurement of a single electron spin (LD qubit) via a readout ancilla based on a singlet-triplet qubit (ST qubit)\cite{Petta:2005jw} in a GaAs/AlGaAs triple quantum dot (TQD) device (Fig.~\ref{fig:system}a). The two states of the electron spin split by the Zeeman energy $E_\text{Z}$, $\ket{\sigma}=\ket{\uparrow}$ or $\ket{\downarrow}$, serve as a natural basis of the LD qubit, while the ancilla ST qubit is encoded in a two-spin subspace of $\ket{\uparrow\downarrow}$ and $\ket{\downarrow\uparrow}$ split by the Zeeman field gradient $\Delta E_\text{Z}$ between the center and the right dots (see Methods for the device design and setup).
This system allows us to extract the information on the single spin state by rapidly measuring the ancilla state\cite{Barthel2009} after entangling the two by a controlled-$Z$ rotation\cite{Mehl2015a,Noiri2017a}(Fig.~\ref{fig:system}b).
The QND nature of the protocol\cite{Ralph2006} is demonstrated by a monotonic increase of the readout fidelity in repeated ancilla measurements. We observe quantum jumps of the single spin dominated by spontaneous relaxation and thermal excitation, further demonstrating very small measurement-induced disturbance.

The experiment is performed by repeating the sequence shown in Fig.~\ref{fig:system}c. Each sequence begins with the preparation of the single spin, followed by the QND readout cycles indexed by $k$, and finishes with a destructive readout.
In the preparation step, the single spin is initialized to $\ket{\uparrow}$ by the energy-selective tunneling\cite{Elzerman2004} and coherently driven by the micromagnet electron spin resonance (MM-ESR)\cite{Tokura:2006ir,Yoneda2014}. The microwave burst duration $\tau_\text{mw}$ is chosen to adjust the expectation value $\left<\hat{\sigma}_z(t=0;\tau_\text{mw})\right>$ of the observable $\hat{\sigma}_z$ for the $z$ spin component. Its eigenvalue $\sigma_z=+1$ ($-1$) corresponds to the $\ket{\sigma}=\ket{\uparrow}$ ground state ($\ket{\downarrow}$ excited state).
The $k$-th QND readout `cycle' is performed at time $t=t_k$ to infer $\sigma_z(t_k)$, the value of $\hat{\sigma}_z$. The ancilla is initialized to the singlet state $\ket{\text{S}}$ (an eigenstate of $\hat{\sigma}_x^\text{ST}$) followed by a controlled-$Z$ rotation\cite{Noiri2017a} with a spin-dependent angle proportional to the interaction time $\tau_k$. The ancilla is then projectively measured in the singlet-triplet basis, resulting in the outcome $M_k \in \{\text{S},\,\text{T}\}$. This process correlates $\sigma_z(t_k)$ and $M_k$, allowing us to infer $\sigma_z(t_k)$ to be $m_k$ as described below. [$m_k=\pm 1$ is called an estimator for the unknown value of $\sigma_z(t_k)$.] The QND readout cycle is consecutively repeated for $k=1,2,\cdots,100$, varying $\tau_k$ as $\tau_k = k \times 0.83\,\text{ns}$.
A hundred consecutive cycles, each of which takes $7\,\mu\text{s}$ to perform, constitutes a `record'.
Finally, the sequence is finished by a destructive measurement of the single spin\cite{Elzerman2004}, with an outcome denoted by $\sigma_z(t=700\,\mu\text{s};\tau_\text{mw})=m^\text{L}$.
The whole sequence is run $50$ times with $\tau_\text{mw}$ varied from $10\,\text{ns}$ to $500\,\text{ns}$. The block of these $50$ sequences is then repeated $800$ times.

\begin{figure}
\begin{centering}
\includegraphics[width=1\textwidth]{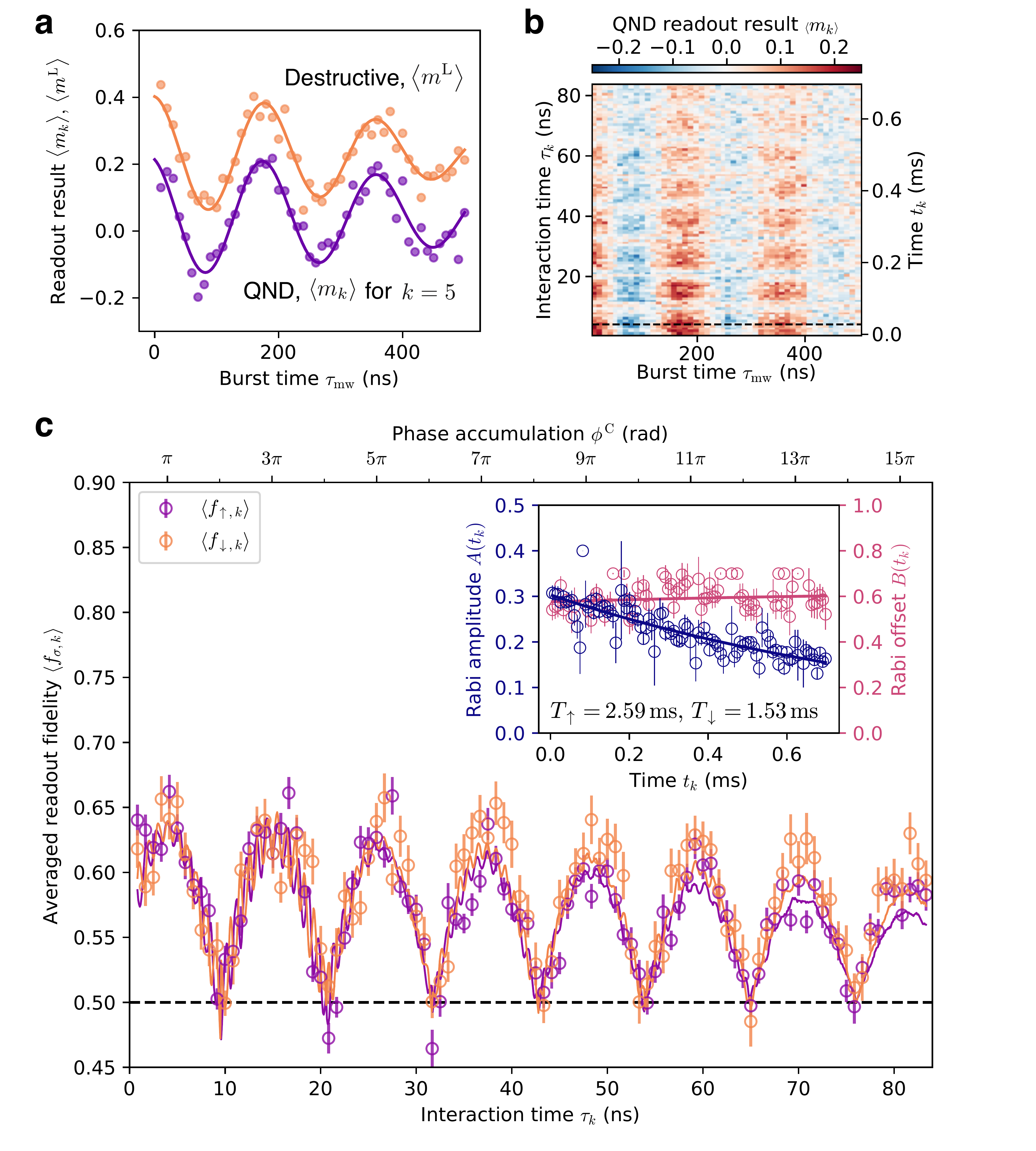}
\par\end{centering}

\protect\caption{\textbf{Demonstration of QND measurement and fidelity analysis.}
\textbf{a,}
Averaged QND readout estimators $\left<m_k\right>$ (taken at $k=5,\,\tau_k=4.2\,\text{ns}$ as an example, where $\phi^\text{C}$ is close to $\pi$) and destructive readout outcomes $\left<m^\text{L}\right>$, plotted against the microwave burst time $\tau_\text{mw}$ for the MM-ESR. 
\textbf{b,}
Averaged QND readout estimators versus the burst time $\tau_\text{mw}$ and the cycle index $k$.
Both the interaction time $\tau_k = k \times 0.83\,\text{ns}$ (left axis) and the laboratory time $t_k = (k-1) \times 7\,\mu\text{s}$ (right axis) are specified by $k$. The horizontal black dashed line indicates $k=5$ shown in \textbf{a}.
\textbf{c,}
Averaged readout fidelities $\left<f_{\sigma,k}\right>$ for the spin states in $\sigma_z(t_k)=\pm 1$ (purple and orange circles), extracted from the analysis of the joint probabilities described in Methods. Error bars represent standard errors obtained from the least mean squares method. The solid curves show numerically simulated fidelities, with their envelopes decaying with different dephasing times ($T_{2\uparrow}^{*}=177\,\text{ns}$ and $T_{2\downarrow}^{*}=212\,\text{ns}$, see Methods) of the ancilla qubit for $\sigma_z=+1$ and $-1$.
The inset shows the amplitude $A(t_k)$ (left axis) and offset $B(t_k)$ (right axis) of the actual qubit oscillation, decaying with time $t_k$. The solid curves are the fits with the exponential damping, allowing us to extract the spin relaxation times $T_\uparrow=2.46\pm0.10\,\text{ms}$ and $T_\downarrow=1.53\pm0.07\,\text{ms}$ for $\ket{\uparrow}$ and $\ket{\downarrow}$ initial states, respectively (see Methods).
}
\label{fig:fidelity}
\end{figure}

In each QND measurement, we assign the spin $\sigma_z(t_k)$ to be $m_k$ ($=\pm 1$) if the conditional probabilities $P(\sigma_z|M_k)$ for a given ancilla measurement outcome $M_k$ satisfy $P(m_k|M_k) > P(-m_k|M_k)$.
This inequality is calculated using the Bayes' theorem as $\frac{P(m_k|M_k)}{P(-m_k|M_k)}=\frac{P(M_k|m_k)}{P(M_k|-m_k)}$ where $P(M_k|\sigma_z)$ is the \emph{likelihood} of finding an ancilla outcome $M_k$ for a given eigenvalue of the input $\sigma_z$.
From an \emph{a priori} characterization of the controlled-$Z$ rotation\cite{Noiri2017a}, $P(M_k|\sigma_z)$ is found as
\begin{equation}
P(M_k=\text{S}|\sigma_z) = 1 - P(M_k=\text{T}|\sigma_z) = a \cos(\phi_{\sigma}+\phi^A) \exp[-(\tau_{k}/T_{2}^{*})^{2}] + b ,
\label{eq:Ps}
\end{equation}
where $\phi_{\sigma}=-\frac{J}{2\hbar}(\tau_k+\tau_0)\sigma_z$ is the phase conditioned on $\sigma_z$, through which the single spin and the ancilla qubit are entangled. Here, $J$ is the inter-qubit exchange coupling and $\tau_0$ is the effective switching time of $J$. The unconditional part $\phi^\text{A}(\tau_k)$ represents the phase of a non-interacting ancilla qubit and $T_{2}^{*}$ is the dephasing time of the ancilla within a single record\cite{Delbecq2015}. Imperfections of the protocol such as the state preparation and measurement errors of the ancilla qubit, tilt of the qubit rotation axis during the controlled-$Z$ rotation, and leakage to non-qubit states are parameterized together by $a$ and $b$. The values of all these parameters except $\phi^\text{A}$ are determined from the measured data by maximum-likelihood estimation prior to the QND measurement (see Methods for details).
The value of $\phi^\text{A}$ in Eq.~\eqref{eq:Ps}, which drifts randomly due to magnetic and charge noises throughout the experiment, is continuously monitored by the Bayesian inference and updated before every execution of the sequence\cite{Noiri2017a,Delbecq2015}.
Figure~\ref{fig:fidelity}a shows the estimator $\left<m_k\right>$ taken at $k=5$ and the destructive readout result $\left<m^\text{L}\right>$ as functions of the microwave burst time $\tau_\text{mw}$, ensemble-averaged over the blocks. Both measurement outcomes exhibit clear Rabi oscillations of the single spin. The QND readout estimators $\left<m_k\right>$ taken at different cycles are plotted in Fig.~\ref{fig:fidelity}b, showing that the visibility of the oscillations varies with $k$ because the degree of the entanglement changes with $\tau_k$.

An essential figure of merit in the QND readout is the fidelity, which is the probability of obtaining a correct estimator $m_k$ when a qubit with a known eigenvalue of $\sigma_z(t_k)$ is given at the time of measurement $t_k$.
Evaluation of the fidelity in this strict sense is, however, often impractical because it apparently demands a perfect preparation of the input state.
We separate the state preparation error by analyzing the joint probabilities of the QND and destructive readouts for the same input states (see Methods for the detailed procedure). Figure \ref{fig:fidelity}c shows the extracted QND and destructive readout fidelities as well as the state preparation error parameterized by the amplitude $A(t_k)$ and offset $B(t_k)$ of the actual Rabi oscillation of $\left<\sigma_z(t_k;\tau_\text{mw})\right>$.
The QND readout fidelities $\left< f_{\uparrow,k} \right>$ and $\left< f_{\downarrow,k} \right>$ for up and down spin states show damped oscillations reflecting the accumulation of the controlled phase during the interaction; they reach maxima (minima) when $\phi^\text{C}=\phi_\downarrow-\phi_\uparrow$ is an odd (even) multiple of $\pi$. 
Those extracted fidelity values agree very well with the numerical simulation (see Methods) plotted as the solid curves.
The spin relaxation times extracted from the exponential decays of $A(t_k)$ and $B(t_k)$ suggest that the possible disturbance due to the readout protocol is small as discussed later in detail. This is a key feature of the QND measurement that allows one to repeat the measurement of an observable to enhance the readout fidelity.

\begin{figure}
\begin{centering}
\includegraphics[width=1\textwidth]{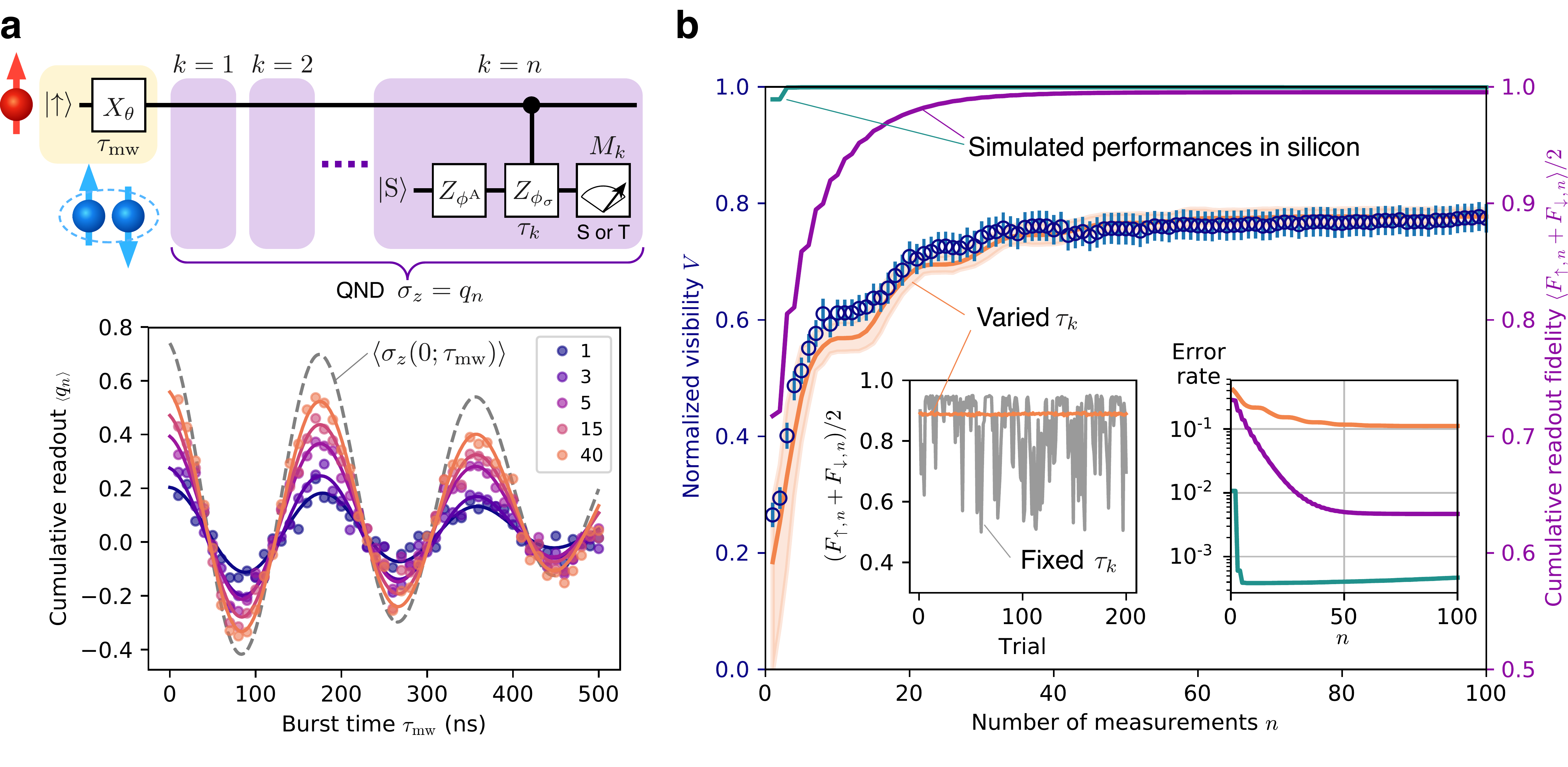}
\par\end{centering}

\protect\caption{\textbf{Fidelity boost in repetitive QND readouts.}
\textbf{a,}
Visibility enhancement of single-spin Rabi oscillations with repeated QND readouts. The upper panel shows the sequence for preparing an input state and obtaining an estimator $q_n$ for $\sigma_z(0;\tau_\text{mw})$ from the QND readout cycles repeated for $n$ times with varied interaction time $\tau_k$. The visibility of the Rabi oscillation increases and approaches the actual expectation value $\left<\sigma_z(0;\tau_\text{mw})\right>$ as the readout fidelity is increased with $n$ (lower panel).
\textbf{b,}
Normalized visibility of the Rabi oscillations in \textbf{a} (blue circles, left axis) plotted against the number of consecutive measurements $n$. The normalized visibility is defined as $V_n=A_n/A(0)$, where $A_n$ is the amplitude of the measured Rabi oscillation in \textbf{a} and $A(0)=0.298$ is the amplitude of the actual qubit oscillation at $t\rightarrow 0$ found in Fig.~\ref{fig:fidelity}c. This normalized visibility relates to the averaged fidelity as $\left<F_{\uparrow,n} + F_{\downarrow,n}\right> / 2 = (V_n + 1) / 2$ shown on the right axis. 
The orange curve represents the simulated fidelity, with the orange shaded region showing the variation of the simulated fidelity (bounded by its minima and maxima) due to the drift of $\phi^\text{A}$.
The purple and green curves show the fidelities simulated for silicon quantum dots with parameters given in the main text.
The left inset shows the variations of the fidelities with a fixed interaction time of $\tau_k = 3.88\,\text{ns}$ (grey) and with varied $\tau_k$ (orange), simulated for $200$ trials with random samples of $\phi^\text{A}$.
The right inset shows the error rate $1 - \left<F_{\uparrow,n} + F_{\downarrow,n}\right>/2$ in each case.
The readout error of the green curve slightly increases with $n>8$, which is an artifact of the approximation neglecting multiple spin flips in the spin trajectories in Eq.~\eqref{eq:Pcum}.
}
\label{fig:repeat}
\end{figure}

To demonstrate this potential, we use a set of measurement outcomes $\{M_k\}$ obtained from $n$ consecutive QND readout cycles to calculate a single \emph{cumulative estimator}, $q_n$.
The probability of the spin being initially in a state with $\sigma_z(0;\tau_\text{mw})=\sigma_0$ is given by $P(\sigma_0|\{M_k\}) \propto P(\{M_k\}|\sigma_0)$ with
\begin{equation}
P(\{M_k\}|\sigma_0) = \sum_{\{\sigma_k\}} \left[\prod_{i=1}^{n} P(M_i|\sigma_i) P(\sigma_i|\sigma_{i-1})\right]
\label{eq:Pcum}
\end{equation}
where $\sigma_k=\sigma_z(t_k;\tau_\text{mw})$ and the sum is taken over all possible realizations of the spin trajectories $\{\sigma_k\}$. This is the optimal estimation exploiting all the available information\cite{Gambetta2007}. Since the spin lifetimes exceed the total measurement time $t_n$, trajectories involving multiple spin flips are rarely realized and therefore neglected in the analysis below. Using the state transfer probability $P(\sigma_i|\sigma_{i-1})$ calculated from the rate equation (see Eq.~\eqref{eq:rate} in Methods), we obtain an estimator $q_n$ again by imposing $P(q_n|\{M_k\}) > P(-q_n|\{M_k\})$.
Figure~\ref{fig:repeat}a shows the visibility improvement of the Rabi oscillations with increasing $n$. The extracted visibility is plotted in Fig.~\ref{fig:repeat}b as a function of $n$ together with the numerical simulation of the averaged fidelity $\left<F_{\uparrow,n}+F_{\downarrow,n}\right>/2$ shown by the orange curve (see Methods for the fidelity derivation). We find monotonic increase of the fidelity up to $0.89$ with $n=100$. We do not see noticeable increase of the fidelity at $n \gtrsim 60$, with its upper bound mainly imposed by the spin relaxation. Indeed, one can no longer gain information from the readout outcomes at times when the spin becomes decorrelated with its initial state (see Supplementary Material for the explicit evaluation of the correlation).

One would expect the best cumulative readout fidelity $F_{\sigma,n}$ by repeating the readout cycles with $\tau_k$ fixed at an optimal value such that the single readout fidelity $\left<f_{\sigma,k}\right>$ is maximal. However, this would lead to the fluctuation of $F_{\sigma,n}$ as plotted in the left inset of Fig.~\ref{fig:repeat}b. The reason is that the total phase of the ancilla qubit fluctuates record-by-record with the drift of $\phi^\text{A}$, so that one cannot distinguish the spin state when $P(M_k|\sigma_z=+1)\approx P(M_k|\sigma_z=-1)$.
The fidelity $F_{\sigma,n}$ can be made robust against the drift of $\phi^\text{A}$ by sampling $\{M_k\}$ with varied values of $\tau_k$ in a set of readout cycles as shown in Fig.~\ref{fig:repeat}b.
On the other hand, the repetitive measurement with an optimal $\tau_k$ would be feasible in materials with less magnetic noise such as silicon. Since the spin relaxation time also tends to be longer in those materials, the QND readout fidelity will be boosted significantly. The purple curve in Fig.~\ref{fig:repeat}b shows the fidelity estimated for a natural silicon quantum dot\cite{Takeda2016,Zajac2017} with $T_2^{*}=1.84\,\mu\text{s}$ and $T_\uparrow=22\,\text{ms}$ ($T_\downarrow=35\,\text{ms}$ assuming the same ratio of relaxation times for the ground and excited spin states), suggesting that the fidelity reaches $99.5\,\%$ at $n=52$. With a better readout visibility of $98\,\%$ reported for the ST qubit\cite{Eng2015}, it even reaches $99.96\,\%$ at $n=5$ well beyond the fault-tolerant threshold\cite{Martinis2015} as shown by the green curve.

\begin{figure}
\begin{centering}
\includegraphics[width=1\textwidth]{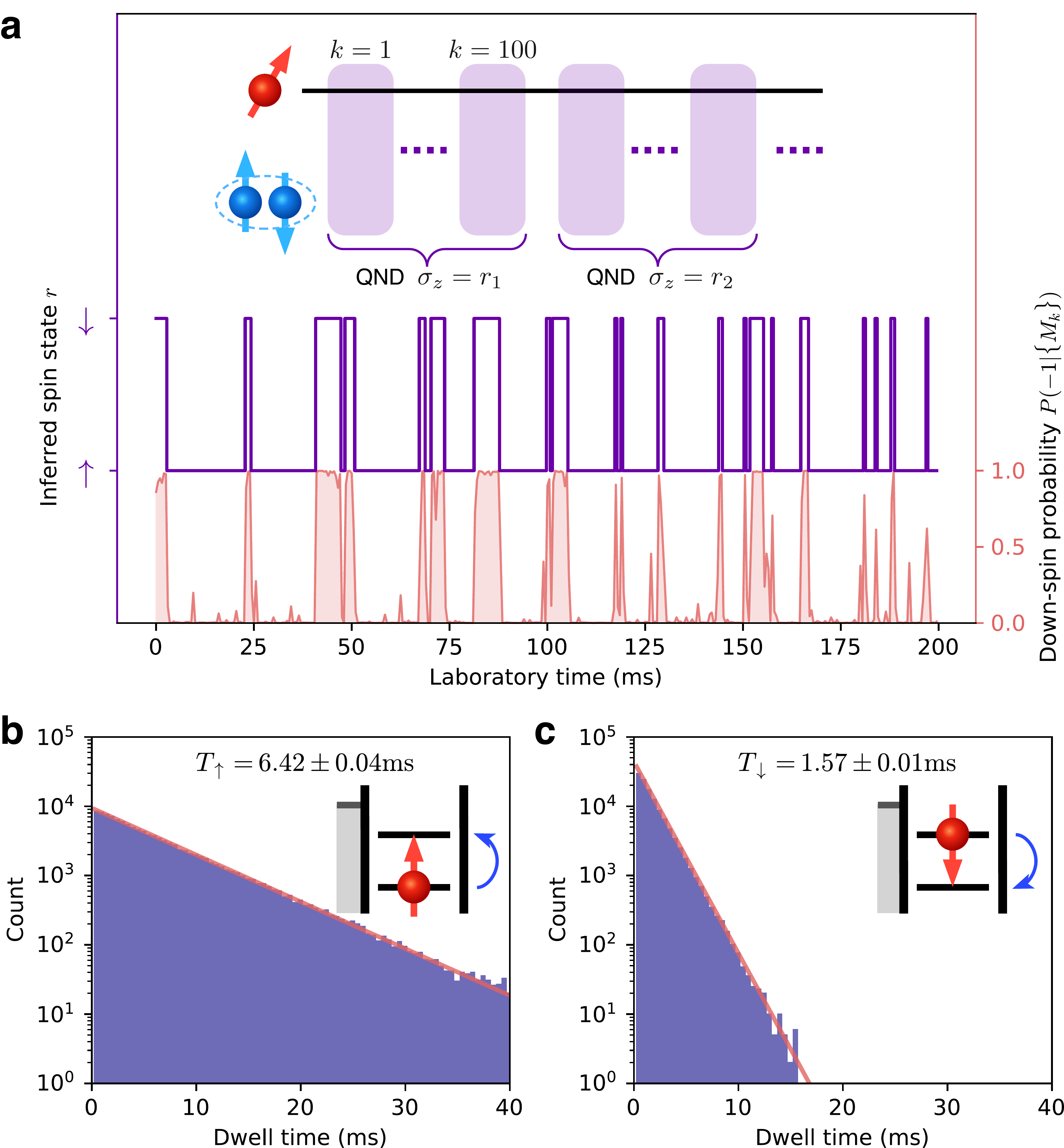}
\par\end{centering}

\protect\caption{\textbf{Quantum jumps of a single electron spin in a quantum dot.}
\textbf{a,}
Realtime dynamics of an electron spin probed by continuous QND readout. The experiment is done by skipping the preparation and the destructive readout of the single spin in the sequence shown in Fig.~\ref{fig:system}c and using a QND readout cycle time of $5\,\mu\text{s}$. The gate conditions are slightly changed from those used in Figs.~\ref{fig:fidelity} and \ref{fig:repeat} so that the spin is confined more strongly and the ancilla qubit can be initialized more rapidly.
The spin-down probability $P(\sigma_z=-1|\{M_k\})$ is obtained from $n=100$ readout cycles (see Methods) and plotted in orange. The spin trajectory shown by a purple curve is obtained from cumulative estimators $r_\alpha$ by imposing $P(r_\alpha|\{M_k\})>P(-r_\alpha|\{M_k\})=1-P(r_\alpha|\{M_k\})$.
\textbf{b,c,}
Histograms of the dwell times in up (\textbf{b}) and down (\textbf{c}) spin states. The solid lines are fits to the data with exponential decay with relaxation times $T_{\uparrow,\downarrow}$.
}
\label{fig:quantumjump}
\end{figure}

Finally, we demonstrate that we can follow the dynamics of an isolated electron spin in a quantum dot\cite{Vamivakas2010}. Figure~\ref{fig:quantumjump} shows spontaneous spin-flip events continuously monitored by cumulative estimators for $n=100$. Here the TQD gate conditions are adjusted to make a stronger confinement potential for the single spin and to suppress possible electron exchange with the reservoir. The statistics of the dwell times, acquired during the total acquisition time of $1000\,\text{s}$, show relaxation times $T_\uparrow=6.42\,\text{ms}$ and $T_\downarrow=1.57\,\text{ms}$ for up and down spin states. Those values give an upper bound of the measurement-induced spin-flip rate of $0.3\,\%$ per cycle (or $27\,\%$ per record), which could be caused by, e.g., the state leakage or the spin-electric coupling to the measurement pulse. Those disturbances would, however, perturb the spin states randomly leading to an expectation $T_\uparrow \approx T_\downarrow$. Since we do not observe such relation, we conclude that the excited-state lifetime $T_\downarrow$ is most probably dominated by the spin-environment coupling rather than the direct measurement disturbance.
Indeed, the $T_\downarrow$ value is in line with the theoretical prediction\cite{Tokura:2006ir} taking into account the large slanting Zeeman field of $>0.6\,\text{T}/\mu\text{m}$ due to the micromagnet, although shorter than those reported for devices without micromagnets\cite{Amasha2008a,Baart2016a}.
Regarding the spin as a two-level system weakly coupled to a bath in thermal equilibrium with $T_\downarrow/T_\uparrow=\exp(-E_\text{Z}/kT_\text{B})$, we find the bath temperature $T_\text{B} \approx 0.5\,\text{K}$ significantly higher than the electron temperature $T_\text{e}\approx 120\,\text{mK}$ measured by Coulomb blockade. This level of heating is reasonable because we observe that the electron temperature increases as the repetition frequency of the pulse for the QND protocol is increased. Heating could be reduced by either reducing the frequency or by increasing the dot-to-gate capacitive coupling so that the pulse amplitude can be decreased.
Irrespective of these further precautions, the value of $T_\downarrow$ is almost unaffected by the protocol, evidencing the QND-ness of our measurement: the evolution of the measured observable is perturbed negligibly by the back action of the measurement or by undesired interactions\cite{Imoto1985,Ralph2006}.

To summarize, we have implemented quantum nondemolition measurement of a single-electron spin qubit via an ancillary singlet-triplet qubit in an array of GaAs gated quantum dots.
The fast and non-invasive readout of the single electron spin demonstrated here brings measurement-based quantum information processing protocols within experimental reach, opening a promising route towards quantum error correction.
We conclude that the application of this technique to silicon spin qubits will enable qubit readout with high fidelity, well beyond the fault-tolerant threshold.

\begin{methods}

\section*{Device design and setup}
The TQD is fabricated on an epitaxially-grown GaAs/AlGaAs heterostructure wafer with a two-dimensional electron gas $100\,\text{nm}$ below the surface.
The Ti/Au gate electrodes deposited on top of the wafer are negatively biased to confine single electrons in each of the TQD and to define the charge sensing quantum dot.
The Co micromagnet is directly placed on the surface and magnetized by the in-plane magnetic field of $B_\text{ext}=3.155\,\text{T}$. It is designed to provide the local Zeeman field difference of about $60\,\text{mT}$ ($40\,\text{mT}$) between the left and center (the center and right) dots as well as the slanting magnetic field necessary for the selective MM-ESR. At the same time, the Zeeman field difference leads to the $\ket{\uparrow\downarrow}$ and $\ket{\downarrow\uparrow}$ eigenstates of the ST qubit which are split from the spin-polarized triplet states by $B_\text{ext}$. The experiment was conducted in a dilution refrigerator and the electron temperature was measured to be about $120\,\text{mK}$.

The initialization, manipulation and destructive readout of the single spin are performed within the $(N_\text{L}, N_\text{C}, N_\text{R})=(1,0,2)$ charge configuration, where $N_\text{L}$ ($N_\text{C}$, $N_\text{R}$) is the number of electrons in the left (center, right) quantum dot. The spin is initialized to the up-spin ground state by exchanging electrons with the reservoir, manipulated by the MM-ESR, and read out by the energy-selective tunneling to the reservoir.
The ST qubit is also initialized to the doubly-occupied singlet state by exchanging electrons with the reservoir near the boundary between $(1,0,2)$ and $(1,0,1)$. Then the singlet is brought to $(1,1,1)$ from $(1,0,2)$ by the rapid adiabatic passage\cite{Taylor2007} and the exchange coupling to the single-spin qubit is turned on near the $(1,1,1)$-$(2,0,1)$ charge transition. The ST qubit is read out by bringing the system back to the $(1,0,2)$ region and detecting whether the double occupancy of the right dot is realized or not. If the measured charge state is $(1,0,2)$ we find the final state to be $M_k=\text{S}$ while the final state is found to be $M_k=\text{T}$ if the system remains in the $(1,1,1)$ charge state.
More details of the device characterization and measurement schemes are given in Ref.~\citen{Noiri2017a}.

\section*{Probability of finding singlet outcomes}
The probability of finding a singlet outcome $M_k=\text{S}$ conditioned on the single-spin (LD qubit) state is ideally given by\cite{Noiri2017a} $P(M_k=\text{S}|\sigma_z)=\bracketii{\sigma \text{S}}{(Z^\text{LD}(-\phi^\text{LD}+\phi_\downarrow)\otimes Z^\text{ST}(-\phi^\text{A}-\phi_\uparrow))CZ(-\phi^\text{C})}{\sigma \text{S}}$, where $Z^\text{LD}$ and $Z^\text{ST}$ are the phase gates for the LD and ST qubits and $CZ(\varphi)$ is the controlled-$Z$ gate with an arbitrary rotation with phase $\varphi$. Here the LD qubit phase is given by $\phi^\text{LD}=E_\text{Z}(\tau_k+\tau_\text{R})/\hbar$ with $\tau_\text{R}=24\,\text{ns}$ the total ramp time of the voltage step used when turning on the exchange interaction $J$. Imperfections of the measurement and dephasing lead to the expression in Eq.~\eqref{eq:Ps}. It comprises a number of parameters unchanged during the experiment and a few parameters varying with time. The latter includes the LD qubit state $\sigma_z$, the Zeeman energy of the ST qubit $\Delta E_\text{Z}$ drifting with the nuclear spin diffusion or the charge noise, and the initial ST qubit phase $\phi$ which accumulates during the loading process from a doubly-occupied singlet in the right-most quantum dot. Here $\Delta E_\text{Z}$ and $\phi$ contribute to $\phi^\text{A}$ via $\phi^\text{A}=\frac{\Delta E_\text{Z}}{\hbar}(\tau_k+\tau_\text{R})+\phi$.
Without requiring knowledge of the trajectories of those varying parameters in the data set, the values of the unchanged parameters are determined from maximum-likelihood estimators by marginalizing out $\sigma_z$, $\Delta E_\text{Z}$ and $\phi$. In this way, we find $J=90.4\pm0.3\,\text{MHz}$, $\tau_0=1.54\pm0.17\,\text{ns}$, $a=0.218\pm0.005$ and $b=0.511\pm0.003$. We also find that the value of $T_2^{*}$ is dependent on the spin state such that $T_{2\uparrow}^{*}=177\pm29\,\text{ns}$ for $\sigma_z=+1$ and $T_{2\downarrow}^{*}=212\pm57\,\text{ns}$ for $\sigma_z=-1$ (see Supplementary Material for the origin of the difference).
Once the values of the constant parameters are specified, the drift of $\Delta E_\text{Z}$ and $\phi$ is continuously monitored by the Bayesian inference, and then $P(M_k|\sigma_z)$ in Eq.~\eqref{eq:Ps} is updated from every record preceding each readout sequence. Reference~\citen{Noiri2017a} contains more details of this procedure.

For the data in Fig.~\ref{fig:quantumjump}, where the gate bias condition is slightly changed, the parameter values are re-estimated to be $J=84.3\pm0.1\,\text{MHz}$, $\tau_0=1.14\pm0.06\,\text{ns}$, $a=0.2180\pm0.0015$ and $b=0.5009\pm0.0014$ ($T_{2\uparrow}^{*}$ and $T_{2\downarrow}^{*}$ are assumed to be unchanged).

\section*{Evolution of the single electron spin}
In the experimental sequence shown in Fig.~\ref{fig:system}c, the LD qubit state is initially prepared by the microwave burst of duration $\tau_\text{mw}$ and then freely evolves with $t$.
The spin-up probability of the LD qubit is then written as $p_\uparrow(t;\tau_\text{mw}) = (1 + \left<\sigma_z(t;\tau_\text{mw})\right>)/2 = A(t) e^{-(\tau_\text{mw}/T_2^\text{Rabi})^2} \cos(2\pi f_\text{Rabi} \tau_\text{mw} + \varphi) + B(t)$, where the amplitude $A(t)$ and the offset $B(t)$ of the Rabi oscillation decay due to the spin relaxation. From fits such as those in Fig.~\ref{fig:fidelity}a, we find the Rabi frequency $f_\text{Rabi} = 5.46\pm0.04\,\text{MHz}$, $\varphi=0.264\pm0.056$ and the decay time $T_2^\text{Rabi} = 526\pm42\,\text{ns}$.

We assume that the evolution of $p_\uparrow(t;\tau_\text{mw})$ follows the rate equation
\begin{equation}
\frac{dp_\uparrow(t;\tau_\text{mw})}{dt} = -\frac{1}{T_\uparrow}p_\uparrow(t;\tau_\text{mw})+\frac{1}{T_\downarrow}(1 - p_\uparrow(t;\tau_\text{mw})) ,
\label{eq:rate}
\end{equation}
where $T_{\uparrow(\downarrow)}$ is the lifetime of the up (down) spin state.
This leads to the exponential decay of $A(t)$ and $B(t)$ such that $A(t)=A(0)e^{-t/T_1}$ and $B(t)=[B(0)-T_1/T_\downarrow]e^{-t/T_1}+T_1/T_\downarrow$ with $T_1^{-1}=T_\uparrow^{-1}+T_\downarrow^{-1}$.
We can thus derive the values of $T_\uparrow$, $T_\downarrow$, $A(0)=0.298\pm0.004$ and $B(0)=0.616\pm0.041$ from the fitting to the data on $A(t_k)$ and $B(t_k)$.

Equation~\eqref{eq:rate} also gives the qubit state transfer probability $P(\sigma_i|\sigma_{i-1})$ used in Eq.~\eqref{eq:Pcum} which describes the probability of flipping the spin state from $\sigma_{i-1}$ to $\sigma_i$ between each measurement cycle. Namely, we obtain $P(+1|+1)=e^{-\Delta t/T_\uparrow}$, $P(-1|-1)=e^{-\Delta t/T_\downarrow}$, $P(-1|+1)=1-e^{-\Delta t/T_\uparrow}$ and $P(+1|-1)=1-e^{-\Delta t/T_\downarrow}$ with $\Delta t = 7\,\mu\text{s}$. Note that we define the initial qubit state to be $\sigma_0=\sigma_1$, i.e., $P(\sigma_1|\sigma_0)=\delta_{\sigma_1 \sigma_0}$.

\section*{Extraction of readout fidelities from joint probabilities}
We introduce fidelities $f_{\sigma,k}$ and $f^\text{L}_{\sigma,k}$ to denote the probabilities of measuring the spin state prepared in the $k$-th cycle $\sigma_z(t_k)$ correctly by the $k$-th QND readout and the final destructive readout, respectively (see Extended Data Fig.~\ref{fig:jointprobability}).
Here both $f_{\sigma,k}$ and $f^\text{L}_{\sigma,k}$ depend on index $k$, because $f_{\sigma,k}$ is a function of the interaction time $\tau_k$, while $f^\text{L}_{\sigma,k}$ is influenced by the spin relaxation taking place during the time interval of $(101-k) \times 7\,\mu\text{s}$ before the destructive readout (see Fig.~\ref{fig:system}c).
The joint probabilities $P_{m_k \cap m^\text{L}}$ of finding an estimator $m_k$ in the $k$-th QND readout and an outcome $m^\text{L}$ in the destructive readout are given by
\begin{align*}
P_{+1 \cap +1}(t_k) &= f_{\uparrow,k} f^\text{L}_{\uparrow,k} p_{\uparrow}(t_k) + (1 - f_{\downarrow,k}) (1 - f^\text{L}_{\downarrow,k}) (1 - p_{\uparrow}(t_k)), \\
P_{+1 \cap -1}(t_k) &= f_{\uparrow,k} (1 - f^\text{L}_{\uparrow,k}) p_{\uparrow}(t_k) + (1 - f_{\downarrow,k}) f^\text{L}_{\downarrow,k} (1 - p_{\uparrow}(t_k)), \\
P_{-1 \cap +1}(t_k) &= (1 - f_{\uparrow,k}) f^\text{L}_{\uparrow,k} p_{\uparrow}(t_k) + f_{\downarrow,k} (1 - f^\text{L}_{\downarrow,k}) (1 - p_{\uparrow}(t_k)), \\
P_{-1 \cap -1}(t_k) &= (1 - f_{\uparrow,k}) (1 - f^\text{L}_{\uparrow,k}) p_{\uparrow}(t_k) + f_{\downarrow,k} f^\text{L}_{\downarrow,k} (1 - p_{\uparrow}(t_k)).
\end{align*}
Note that we use the fact that the QND readout and the destructive readout are perfomed on the same input state in each single-shot sequence.
For each $k$, we find Rabi oscillations of $P_{m_k \cap m^\text{L}}(t_k; \tau_\text{mw})$ as shown in Extended Data Fig.~\ref{fig:jointprobability}.
The correlation of the two readout schemes is clearly seen in the large oscillation amplitudes in the joint probabilities for $m_k=m^\text{L}$ while the anti-correlated signals ($m_k\neq m^\text{L}$) show only small residual oscillations due to readout errors.
By fitting those oscillations, we obtain an overconstrained set of eight equations on $f_{\sigma,k}$, $f^\text{L}_{\sigma,k}$, $A(t_k)$, and $B(t_k)$. We derive the most likely values by the least mean squares method for each $k$ as shown in Fig.~\ref{fig:fidelity}c.

\section*{Theoretical model of readout fidelities}
When we perform a QND measurement and find an outcome $M_k$ from the ancilla readout in the $k$-th cycle, we find a correct estimator $m_k$ for $\sigma_z(t_k)$ only if $P(\sigma_z|M_k) - P(-\sigma_z|M_k) > 0$. The success probability $f_{\sigma,k}$ is given by summing $H(P(\sigma_z|M_k) - P(-\sigma_z|M_k))$ [$H(x)$ is the Heaviside step function] over all possible ancilla readout outcomes $M_k$ realized with probability $P^{\prime}(M_k|\sigma_z)$,
\begin{equation}
f_{\sigma,k} = \sum_{M_k} P^{\prime}(M_k|\sigma_z) H(P(\sigma_z|M_k) - P(-\sigma_z|M_k)) .
\label{eq:fidelity}
\end{equation}
Here, $P^{\prime}(M_k|\sigma_z)$ is approximated by $P(M_k|\sigma_z)$ given by Eq.~\eqref{eq:Ps}, although $P^{\prime}(M_k|\sigma_z)$ may differ from $P(M_k|\sigma_z)$ due to the drifts of $\Delta E_\text{Z}$ and $\phi$ between each cycle indexed by $k$. [Note that $P(M_k|\sigma_z)$ is updated between each record but unchanged between each cycle.] Since the projection angle of the ST qubit against the readout basis changes with $\phi^\text{A}$, the values of $f_{\sigma,k}$ vary over time. Averaging $f_{\sigma,k}$ over fluctuating $\phi^\text{A}$ gives the solid curves of $\left< f_{\sigma,k} \right>$ in Fig.~\ref{fig:fidelity}c, which agrees well with the data.

The fidelity of the cumulative readout using $n$ measurement outcomes is similarly calculated. Equation~\eqref{eq:fidelity} is generalized to
\begin{equation}
F_{\sigma,n} = \sum_{\{M_k\}} P^{\prime}(\{M_k\}|\sigma_0) H(P(\sigma_0|\{M_k\}) - P(-\sigma_0|\{M_k\})) .
\label{eq:repeat}
\end{equation}
Here, $P^{\prime}(\{M_k\}|\sigma_0)$ is similar to $P(\{M_k\}|\sigma_0)$ in Eq.~\eqref{eq:Pcum}, but additionally taking into account the drifts of $\Delta E_\text{Z}$ and $\phi$ between each cycle indexed by $k$. Thus, rewriting the likelihood in Eq.~\eqref{eq:Ps} as $P_{\Delta E_\text{Z},\phi}(M_k|\sigma_z)$, $P^{\prime}(\{M_k\}|\sigma_0)$ is given by
$$ P^{\prime}(\{M_k\}|\sigma_0) = \sum_{\{\sigma_k,\Delta E_{\text{Z},k}, \phi_k\}} \left[\prod_{i=1}^{n} P_{\Delta E_{\text{Z},i},\phi_i}(M_i|\sigma_i) P(\sigma_i|\sigma_{i-1}) P(\Delta E_{\text{Z},i}|\Delta E_{\text{Z},i-1}) P(\phi_i|\phi_{i-1})\right] .$$
We model the drifts by the Gaussian random walks as $P(\Delta E_{\text{Z},i}|\Delta E_{\text{Z},i-1}) \propto \exp\left[-\frac{(\Delta E_{\text{Z},i}-\Delta E_{\text{Z},i-1})^2}{2\sigma_f^2}\right]$ and $P(\phi_i|\phi_{i-1}) \propto \exp\left[-\frac{(\phi_i-\phi_{i-1})^2}{2\sigma_\phi^2}\right]$ (see Supplementary Material for the values of $\sigma_f$ and $\sigma_\phi$).
The values of $F_{\sigma,n}$ plotted in Fig.~\ref{fig:repeat}b are calculated by simulating $10,000$ numerically generated random sets of outcomes $\{M_k\}$, each corresponding to a random trajectory of $\sigma_k$ following Eq.~\eqref{eq:rate}, $\Delta E_{\text{Z},k}$ and $\phi_k$ following the Gaussian random walks.

\section*{Observation of quantum jumps}

For the data in Fig.~\ref{fig:quantumjump}, each cumulative estimator $r_\alpha$ is obtained imposing $P(r_\alpha|\{M_k\})>P(-r_\alpha|\{M_k\})$ for the $\alpha$-th record of $n=100$ cycles. Using the Bayes theorem, $P(\sigma_\alpha|\{M_k\})$ is given by $P(\sigma_\alpha|\{M_k\})=P(\{M_k\}|\sigma_\alpha) P(\sigma_\alpha) / P(\{M_k\})$. When one has no prior knowledge of $\sigma_\alpha$ [$P(\sigma_\alpha=+1)=P(\sigma_\alpha=-1)=1/2$], the readout fidelity expected in our experiment remains below $0.9$ as discussed in the main text. This imperfect fidelity leads to observation of fake quantum jumps and we find $T_\uparrow$ and $T_\downarrow$ values somewhat smaller than those presented in Fig.~\ref{fig:quantumjump}b,c.

To suppress the readout errors, we use the prior probability distribution $P(\sigma_\alpha)=P(\sigma_\alpha|\sigma_{\alpha-1}) P(\sigma_{\alpha-1})$, where $P(\sigma_\alpha|\sigma_{\alpha-1})$ is the state transfer probability between records and $P(\sigma_{\alpha-1})$ is the probability distribution obtained in the previous record. Here $P(\sigma_\alpha|\sigma_{\alpha-1})$ is given by $P(+1|+1)=e^{-n\Delta t/T_\uparrow}$, $P(-1|-1)=e^{-n\Delta t/T_\downarrow}$, $P(-1|+1)=1-e^{-n\Delta t/T_\uparrow}$ and $P(+1|-1)=1-e^{-n\Delta t/T_\downarrow}$ with $\Delta t = 5\,\mu\text{s}$. We initially use the values of $T_\uparrow$ and $T_\downarrow$ extracted in the above, calculate the spin trajectory, and re-extract the values of $T_\uparrow$ and $T_\downarrow$. After repeating this procedure a few times, we find the values of $T_\uparrow$ and $T_\downarrow$ converge and obtain the result shown in Fig.~\ref{fig:quantumjump}. We tested this procedure in numerical simulations and confirmed that it gives a reliable estimate of $T_\uparrow$ and $T_\downarrow$.

\section*{Data availability}
The data that support the findings of this study are available from the corresponding authors upon reasonable request.

\end{methods}

\begin{addendum}
\item We thank N.~Imoto for fruitful discussions.
We thank RIKEN CEMS Emergent Matter Science Research Support Team and Microwave Research Group at Caltech for technical assistance.
Part of this work was financially supported by
CREST, JST (JPMJCR15N2, JPMJCR1675),
the ImPACT Program of Council for Science, Technology and Innovation (Cabinet Office, Government of Japan),
JSPS KAKENHI Grants No. 26220710, No. JP16H02204, and No. 18H01819.
T.N., T.O., and J.Y. acknowledge financial support from RIKEN Incentive Research Projects.
T.O. acknowledges support from
JSPS KAKENHI Grants No. 16H00817 and No. 17H05187,
PRESTO (JPMJPR16N3), JST,
Yazaki Memorial Foundation for Science and Technology Research Grant,
Advanced Technology Institute Research Grant, the Murata Science Foundation Research Grant,
Izumi Science and Technology Foundation Research Grant, 
TEPCO Memorial Foundation Research Grant, 
The Thermal \& Electric Energy Technology Foundation Research Grant, 
The Telecommunications Advancement Foundation Research Grant, 
Futaba Electronics Memorial Foundation Research Grant, 
and MST Foundation Research Grant.
A.D.W. and A.L. greatfully acknowledge support from Mercur Pr2013-0001,
BMBF Q.Com-H 16KIS0109, TRR160, and DFH/UFA CDFA-05-06.

\item[Author Contributions]
T.N., M.R.D. and S.T. planned the project.
A.L. and A.D.W grew the heterostructure and T.N. and A.N. fabricated the device.
T.N. and A.N. conducted the experiment with the assistance of K.K.;
T.N. and A.N. analyzed the data and wrote the manuscript with the inputs from J.Y. and P.S.
All authors discussed the results and commented on the manuscript.
The project was supervised by S.T.

\item[Author Information]
Reprints and permissions information is available at www.nature.com/reprints.
The authors declare no competing financial interests.
Correspondence should be addressed to T.N. (nakajima.physics@icloud.com) or S.T. (tarucha@ap.t.u-tokyo.ac.jp).
\end{addendum}

\bibliographystyle{naturemag}

\begin{thebibliography}{10}
\expandafter\ifx\csname url\endcsname\relax
  \def\url#1{\texttt{#1}}\fi
\expandafter\ifx\csname urlprefix\endcsname\relax\def\urlprefix{URL }\fi
\providecommand{\bibinfo}[2]{#2}
\providecommand{\eprint}[2][]{\url{#2}}

\bibitem{Grangier1998}
\bibinfo{author}{Grangier, P.}, \bibinfo{author}{Levenson, J.~A.} \&
  \bibinfo{author}{Poizat, J.~P.}
\newblock \bibinfo{title}{{Quantum non-demolition measurements in optics}}.
\newblock \emph{\bibinfo{journal}{Nature}} \textbf{\bibinfo{volume}{396}},
  \bibinfo{pages}{537--542} (\bibinfo{year}{1998}).
\newblock \eprint{arXiv:1011.1669v3}.

\bibitem{Imoto1985}
\bibinfo{author}{Imoto, N.}, \bibinfo{author}{Haus, H. H.~A.} \&
  \bibinfo{author}{Yamamoto, Y.}
\newblock \bibinfo{title}{{Quantum nondemolition measurement of the photon
  number via the optical Kerr effect}}.
\newblock \emph{\bibinfo{journal}{Physical Review A}}
  \textbf{\bibinfo{volume}{32}}, \bibinfo{pages}{2287--2292}
  (\bibinfo{year}{1985}).
\newblock \urlprefix\url{http://link.aps.org/doi/10.1103/PhysRevA.32.2287}.

\bibitem{Mehl2015a}
\bibinfo{author}{Mehl, S.} \& \bibinfo{author}{DiVincenzo, D.~P.}
\newblock \bibinfo{title}{{Simple operation sequences to couple and interchange
  quantum information between spin qubits of different kinds}}.
\newblock \emph{\bibinfo{journal}{Physical Review B}}
  \textbf{\bibinfo{volume}{92}}, \bibinfo{pages}{115448}
  (\bibinfo{year}{2015}).

\bibitem{Noiri2017a}
\bibinfo{author}{Noiri, A.} \emph{et~al.}
\newblock \bibinfo{title}{{A fast quantum interface between different spin
  qubit encodings}}.
\newblock \emph{\bibinfo{journal}{Nature communications}}
  \textbf{\bibinfo{volume}{9}}, \bibinfo{pages}{5066} (\bibinfo{year}{2018}).
\newblock \urlprefix\url{http://arxiv.org/abs/1804.04764}.
\newblock \eprint{1804.04764}.

\bibitem{Lupascu2007}
\bibinfo{author}{Lupascu, A.} \emph{et~al.}
\newblock \bibinfo{title}{{Quantum non-demolition measurement of a
  superconducting two-level system}}.
\newblock \emph{\bibinfo{journal}{Nature Physics}}
  \textbf{\bibinfo{volume}{3}}, \bibinfo{pages}{119--123}
  (\bibinfo{year}{2007}).
\newblock \urlprefix\url{http://dx.doi.org/10.1038/nphys509
  http://www.nature.com/nphys/journal/v3/n2/pdf/nphys509.pdf}.
\newblock \eprint{0611505}.

\bibitem{Jiang2009}
\bibinfo{author}{Jiang, L.} \emph{et~al.}
\newblock \bibinfo{title}{{Repetitive Readout of a Single Electronic Spin via
  Quantum Logic with Nuclear Spin Ancillae}}.
\newblock \emph{\bibinfo{journal}{Science}} \textbf{\bibinfo{volume}{326}},
  \bibinfo{pages}{267--272} (\bibinfo{year}{2009}).
\newblock
  \urlprefix\url{http://www.sciencemag.org/cgi/doi/10.1126/science.1176496}.

\bibitem{Vamivakas2010}
\bibinfo{author}{Vamivakas, A.} \emph{et~al.}
\newblock \bibinfo{title}{{Observation of spin-dependent quantum jumps via
  quantum dot resonance fluorescence.}}
\newblock \emph{\bibinfo{journal}{Nature}} \textbf{\bibinfo{volume}{467}},
  \bibinfo{pages}{297--300} (\bibinfo{year}{2010}).
\newblock
  \urlprefix\url{http://www.nature.com/nature/journal/v467/n7313/abs/nature09359.html}.

\bibitem{Veldhorst2014}
\bibinfo{author}{Veldhorst, M.} \emph{et~al.}
\newblock \bibinfo{title}{{An addressable quantum dot qubit with fault-tolerant
  control-fidelity.}}
\newblock \emph{\bibinfo{journal}{Nature Nanotechnology}}
  \textbf{\bibinfo{volume}{9}}, \bibinfo{pages}{981--985}
  (\bibinfo{year}{2014}).
\newblock \urlprefix\url{http://www.ncbi.nlm.nih.gov/pubmed/25305743}.

\bibitem{Kawakami2014}
\bibinfo{author}{Kawakami, E.} \emph{et~al.}
\newblock \bibinfo{title}{{Electrical control of a long-lived spin qubit in a
  Si/SiGe quantum dot.}}
\newblock \emph{\bibinfo{journal}{Nature Nanotechnology}}
  \textbf{\bibinfo{volume}{9}}, \bibinfo{pages}{666} (\bibinfo{year}{2014}).
\newblock \urlprefix\url{http://www.ncbi.nlm.nih.gov/pubmed/25108810}.

\bibitem{Takeda2016}
\bibinfo{author}{Takeda, K.} \emph{et~al.}
\newblock \bibinfo{title}{{A fault-tolerant addressable spin qubit in a natural
  silicon quantum dot}}.
\newblock \emph{\bibinfo{journal}{Science Advances}}
  \textbf{\bibinfo{volume}{2}}, \bibinfo{pages}{e1600694}
  (\bibinfo{year}{2016}).
\newblock \urlprefix\url{http://arxiv.org/abs/1602.07833}.
\newblock \eprint{1602.07833}.

\bibitem{Yoneda2017}
\bibinfo{author}{Yoneda, J.} \emph{et~al.}
\newblock \bibinfo{title}{{A quantum-dot spin qubit with coherence limited by
  charge noise and fidelity higher than 99.9\%}}.
\newblock \emph{\bibinfo{journal}{Nature nanotechnology}}
  \textbf{\bibinfo{volume}{13}}, \bibinfo{pages}{102--106}
  (\bibinfo{year}{2018}).
\newblock \urlprefix\url{http://arxiv.org/abs/1708.01454}.
\newblock \eprint{1708.01454}.

\bibitem{Veldhorst2015}
\bibinfo{author}{Veldhorst, M.} \emph{et~al.}
\newblock \bibinfo{title}{{A two-qubit logic gate in silicon}}.
\newblock \emph{\bibinfo{journal}{Nature}} \textbf{\bibinfo{volume}{526}},
  \bibinfo{pages}{410--414} (\bibinfo{year}{2015}).
\newblock \urlprefix\url{http://www.nature.com/doifinder/10.1038/nature15263}.

\bibitem{Zajac2017}
\bibinfo{author}{Zajac, D.~M.} \emph{et~al.}
\newblock \bibinfo{title}{{Resonantly driven CNOT gate for electron spins}}.
\newblock \emph{\bibinfo{journal}{Science}} \textbf{\bibinfo{volume}{359}},
  \bibinfo{pages}{439} (\bibinfo{year}{2017}).

\bibitem{Watson2017}
\bibinfo{author}{Watson, T.~F.} \emph{et~al.}
\newblock \bibinfo{title}{{A programmable two-qubit quantum processor in
  silicon}}.
\newblock \emph{\bibinfo{journal}{Nature}} \textbf{\bibinfo{volume}{555}},
  \bibinfo{pages}{633} (\bibinfo{year}{2018}).
\newblock \urlprefix\url{http://arxiv.org/abs/1708.04214}.
\newblock \eprint{1708.04214}.

\bibitem{Nielsen2010}
\bibinfo{author}{Nielsen, M.~A.} \& \bibinfo{author}{Chuang, I.~L.}
\newblock \emph{\bibinfo{title}{{Quantum Computation and Quantum Information}}}
  (\bibinfo{year}{2010}).

\bibitem{Ralph2006}
\bibinfo{author}{Ralph, T.~C.}, \bibinfo{author}{Bartlett, S.~D.},
  \bibinfo{author}{O'Brien, J.~L.}, \bibinfo{author}{Pryde, G.~J.} \&
  \bibinfo{author}{Wiseman, H.~M.}
\newblock \bibinfo{title}{{Quantum nondemolition measurements for quantum
  information}}.
\newblock \emph{\bibinfo{journal}{Phys. Rev. A}} \textbf{\bibinfo{volume}{73}},
  \bibinfo{pages}{012113} (\bibinfo{year}{2006}).
\newblock \eprint{0412149v1}.

\bibitem{Loss1998}
\bibinfo{author}{Loss, D.} \& \bibinfo{author}{DiVincenzo, D.~P.}
\newblock \bibinfo{title}{{Quantum computation with quantum dots}}.
\newblock \emph{\bibinfo{journal}{Physical Review A}}
  \textbf{\bibinfo{volume}{57}}, \bibinfo{pages}{120--126}
  (\bibinfo{year}{1998}).
\newblock \urlprefix\url{http://link.aps.org/doi/10.1103/PhysRevA.57.120}.

\bibitem{Elzerman2004}
\bibinfo{author}{Elzerman, J.~M.} \emph{et~al.}
\newblock \bibinfo{title}{{Single-shot read-out of an individual electron spin
  in a quantum dot}}.
\newblock \emph{\bibinfo{journal}{Nature}} \textbf{\bibinfo{volume}{430}},
  \bibinfo{pages}{431--435} (\bibinfo{year}{2004}).
\newblock
  \urlprefix\url{http://www.nature.com/nature/journal/v430/n6998/abs/nature02693.html}.

\bibitem{Riste2012a}
\bibinfo{author}{Rist{\`{e}}, D.}, \bibinfo{author}{Bultink, C.~C.},
  \bibinfo{author}{Lehnert, K.~W.} \& \bibinfo{author}{Dicarlo, L.}
\newblock \bibinfo{title}{{Feedback control of a solid-state qubit using
  high-fidelity projective measurement}}.
\newblock \emph{\bibinfo{journal}{Phys. Rev. Lett.}}
  \textbf{\bibinfo{volume}{109}}, \bibinfo{pages}{240502}
  (\bibinfo{year}{2012}).
\newblock \eprint{1207.2944}.

\bibitem{Neumann:2010kx}
\bibinfo{author}{Neumann, P.} \emph{et~al.}
\newblock \bibinfo{title}{{Single-Shot Readout of a Single Nuclear Spin}}.
\newblock \emph{\bibinfo{journal}{Science}} \textbf{\bibinfo{volume}{329}},
  \bibinfo{pages}{542--544} (\bibinfo{year}{2010}).

\bibitem{Robledo2011}
\bibinfo{author}{Robledo, L.} \emph{et~al.}
\newblock \bibinfo{title}{{High-fidelity projective read-out of a solid-state
  spin quantum register}}.
\newblock \emph{\bibinfo{journal}{Nature}} \textbf{\bibinfo{volume}{477}},
  \bibinfo{pages}{574--578} (\bibinfo{year}{2011}).
\newblock \urlprefix\url{http://www.nature.com/doifinder/10.1038/nature10401}.
\newblock \eprint{1301.0392v1}.

\bibitem{Pla2013}
\bibinfo{author}{Pla, J.~J.} \emph{et~al.}
\newblock \bibinfo{title}{{High-fidelity readout and control of a nuclear spin
  qubit in silicon}}.
\newblock \emph{\bibinfo{journal}{Nature}} \textbf{\bibinfo{volume}{496}},
  \bibinfo{pages}{334--338} (\bibinfo{year}{2013}).
\newblock \urlprefix\url{http://www.nature.com/doifinder/10.1038/nature12011}.
\newblock \eprint{1302.0047}.

\bibitem{Mi2017}
\bibinfo{author}{Mi, X.} \emph{et~al.}
\newblock \bibinfo{title}{{A Coherent Spin-Photon Interface in Silicon}}.
\newblock \emph{\bibinfo{journal}{Nature}} \textbf{\bibinfo{volume}{555}},
  \bibinfo{pages}{599} (\bibinfo{year}{2017}).
\newblock \urlprefix\url{http://arxiv.org/abs/1710.03265}.
\newblock \eprint{1710.03265}.

\bibitem{Samkharadze2017}
\bibinfo{author}{Samkharadze, N.} \emph{et~al.}
\newblock \bibinfo{title}{{Strong spin-photon coupling in silicon}}.
\newblock \emph{\bibinfo{journal}{Science}} \textbf{\bibinfo{volume}{359}},
  \bibinfo{pages}{1123} (\bibinfo{year}{2017}).
\newblock \urlprefix\url{http://arxiv.org/abs/1711.02040}.
\newblock \eprint{1711.02040}.

\bibitem{Petta:2005jw}
\bibinfo{author}{Petta, J.~R.} \emph{et~al.}
\newblock \bibinfo{title}{{Coherent manipulation of coupled electron spins in
  semiconductor quantum dots}}.
\newblock \emph{\bibinfo{journal}{Science}} \textbf{\bibinfo{volume}{309}},
  \bibinfo{pages}{2180--2184} (\bibinfo{year}{2005}).

\bibitem{Barthel2009}
\bibinfo{author}{Barthel, C.}, \bibinfo{author}{Reilly, D.},
  \bibinfo{author}{Marcus, C.}, \bibinfo{author}{Hanson, M.} \&
  \bibinfo{author}{Gossard, A.}
\newblock \bibinfo{title}{{Rapid Single-Shot Measurement of a Singlet-Triplet
  Qubit}}.
\newblock \emph{\bibinfo{journal}{Phys. Rev. Lett.}}
  \textbf{\bibinfo{volume}{103}}, \bibinfo{pages}{160503}
  (\bibinfo{year}{2009}).
\newblock
  \urlprefix\url{http://link.aps.org/doi/10.1103/PhysRevLett.103.160503}.

\bibitem{Tokura:2006ir}
\bibinfo{author}{Tokura, Y.}, \bibinfo{author}{van~der Wiel, W.~G.},
  \bibinfo{author}{Obata, T.} \& \bibinfo{author}{Tarucha, S.}
\newblock \bibinfo{title}{{Coherent single electron spin control in a slanting
  Zeeman field}}.
\newblock \emph{\bibinfo{journal}{Phys. Rev. Lett.}}
  \textbf{\bibinfo{volume}{96}}, \bibinfo{pages}{47202} (\bibinfo{year}{2006}).

\bibitem{Yoneda2014}
\bibinfo{author}{Yoneda, J.} \emph{et~al.}
\newblock \bibinfo{title}{{Fast Electrical Control of Single Electron Spins in
  Quantum Dots with Vanishing Influence from Nuclear Spins}}.
\newblock \emph{\bibinfo{journal}{Phys. Rev. Lett.}}
  \textbf{\bibinfo{volume}{113}}, \bibinfo{pages}{267601}
  (\bibinfo{year}{2014}).
\newblock
  \urlprefix\url{http://link.aps.org/doi/10.1103/PhysRevLett.113.267601}.

\bibitem{Delbecq2015}
\bibinfo{author}{Delbecq, M.~R.} \emph{et~al.}
\newblock \bibinfo{title}{{Quantum dephasing in a gated GaAs triple quantum dot
  due to non-ergodic noise}}.
\newblock \emph{\bibinfo{journal}{Phys. Rev. Lett.}}
  \textbf{\bibinfo{volume}{116}}, \bibinfo{pages}{046802}
  (\bibinfo{year}{2016}).

\bibitem{Gambetta2007}
\bibinfo{author}{Gambetta, J.}, \bibinfo{author}{Braff, W.},
  \bibinfo{author}{Wallraff, A.}, \bibinfo{author}{Girvin, S.} \&
  \bibinfo{author}{Schoelkopf, R.}
\newblock \bibinfo{title}{{Protocols for optimal readout of qubits using a
  continuous quantum nondemolition measurement}}.
\newblock \emph{\bibinfo{journal}{Physical Review A}}
  \textbf{\bibinfo{volume}{76}}, \bibinfo{pages}{012325}
  (\bibinfo{year}{2007}).
\newblock \urlprefix\url{http://link.aps.org/doi/10.1103/PhysRevA.76.012325}.

\bibitem{Eng2015}
\bibinfo{author}{Eng, K.} \emph{et~al.}
\newblock \bibinfo{title}{{Isotopically enhanced triple-quantum-dot qubit}}.
\newblock \emph{\bibinfo{journal}{Science Advances}}
  \textbf{\bibinfo{volume}{1}}, \bibinfo{pages}{e1500214}
  (\bibinfo{year}{2015}).

\bibitem{Martinis2015}
\bibinfo{author}{Martinis, J.~M.}
\newblock \bibinfo{title}{{Qubit metrology for building a fault-tolerant
  quantum computer}}.
\newblock \emph{\bibinfo{journal}{npj Quantum Information}}
  \textbf{\bibinfo{volume}{1}}, \bibinfo{pages}{15005} (\bibinfo{year}{2015}).
\newblock \urlprefix\url{http://www.nature.com/articles/npjqi20155}.
\newblock \eprint{1510.01406}.

\bibitem{Amasha2008a}
\bibinfo{author}{Amasha, S.} \emph{et~al.}
\newblock \bibinfo{title}{{Electrical control of spin relaxation in a quantum
  dot}}.
\newblock \emph{\bibinfo{journal}{Physical Review Letters}}
  \textbf{\bibinfo{volume}{100}}, \bibinfo{pages}{046803}
  (\bibinfo{year}{2008}).
\newblock \eprint{arXiv:0707.1656v1}.

\bibitem{Baart2016a}
\bibinfo{author}{Baart, T.~A.} \emph{et~al.}
\newblock \bibinfo{title}{{Single-spin CCD}}.
\newblock \emph{\bibinfo{journal}{Nature Nanotechnology}}
  \textbf{\bibinfo{volume}{11}}, \bibinfo{pages}{330--334}
  (\bibinfo{year}{2016}).
\newblock
  \urlprefix\url{http://www.nature.com/doifinder/10.1038/nnano.2015.291}.
\newblock \eprint{1507.07991}.

\bibitem{Taylor2007}
\bibinfo{author}{Taylor, J.} \emph{et~al.}
\newblock \bibinfo{title}{{Relaxation, dephasing, and quantum control of
  electron spins in double quantum dots}}.
\newblock \emph{\bibinfo{journal}{Phys. Rev. B}} \textbf{\bibinfo{volume}{76}},
  \bibinfo{pages}{035315} (\bibinfo{year}{2007}).
\newblock \urlprefix\url{http://link.aps.org/doi/10.1103/PhysRevB.76.035315}.

\bibitem{Reilly2007}
\bibinfo{author}{Reilly, D.~J.}, \bibinfo{author}{Marcus, C.~M.},
  \bibinfo{author}{Hanson, M.~P.} \& \bibinfo{author}{Gossard, A.~C.}
\newblock \bibinfo{title}{{Fast single-charge sensing with a rf quantum point
  contact}}.
\newblock \emph{\bibinfo{journal}{Appl. Phys. Lett.}}
  \textbf{\bibinfo{volume}{91}}, \bibinfo{pages}{162101}
  (\bibinfo{year}{2007}).
\newblock
  \urlprefix\url{http://link.aip.org/link/APPLAB/v91/i16/p162101/s1}.

\bibitem{Barthel:2010fk}
\bibinfo{author}{Barthel, C.} \emph{et~al.}
\newblock \bibinfo{title}{{Fast sensing of double-dot charge arrangement and
  spin state with a radio-frequency sensor quantum dot}}.
\newblock \emph{\bibinfo{journal}{Phys. Rev. B}} \textbf{\bibinfo{volume}{81}},
  \bibinfo{pages}{161308(R)} (\bibinfo{year}{2010}).

\end{thebibliography}

\newpage

\newpage

\newpage

\newpage

\makeatletter
\setcounter{figure}{0}
\renewcommand{\NAT@figcaption}[2][]{%
	\refstepcounter{figure}
	\sffamily\noindent\textbf{Extended Data Figure \arabic{figure}}\hspace{1em}#2
	}
\makeatother

\newpage
\begin{figure}
\begin{centering}
\includegraphics[width=0.8\textwidth]{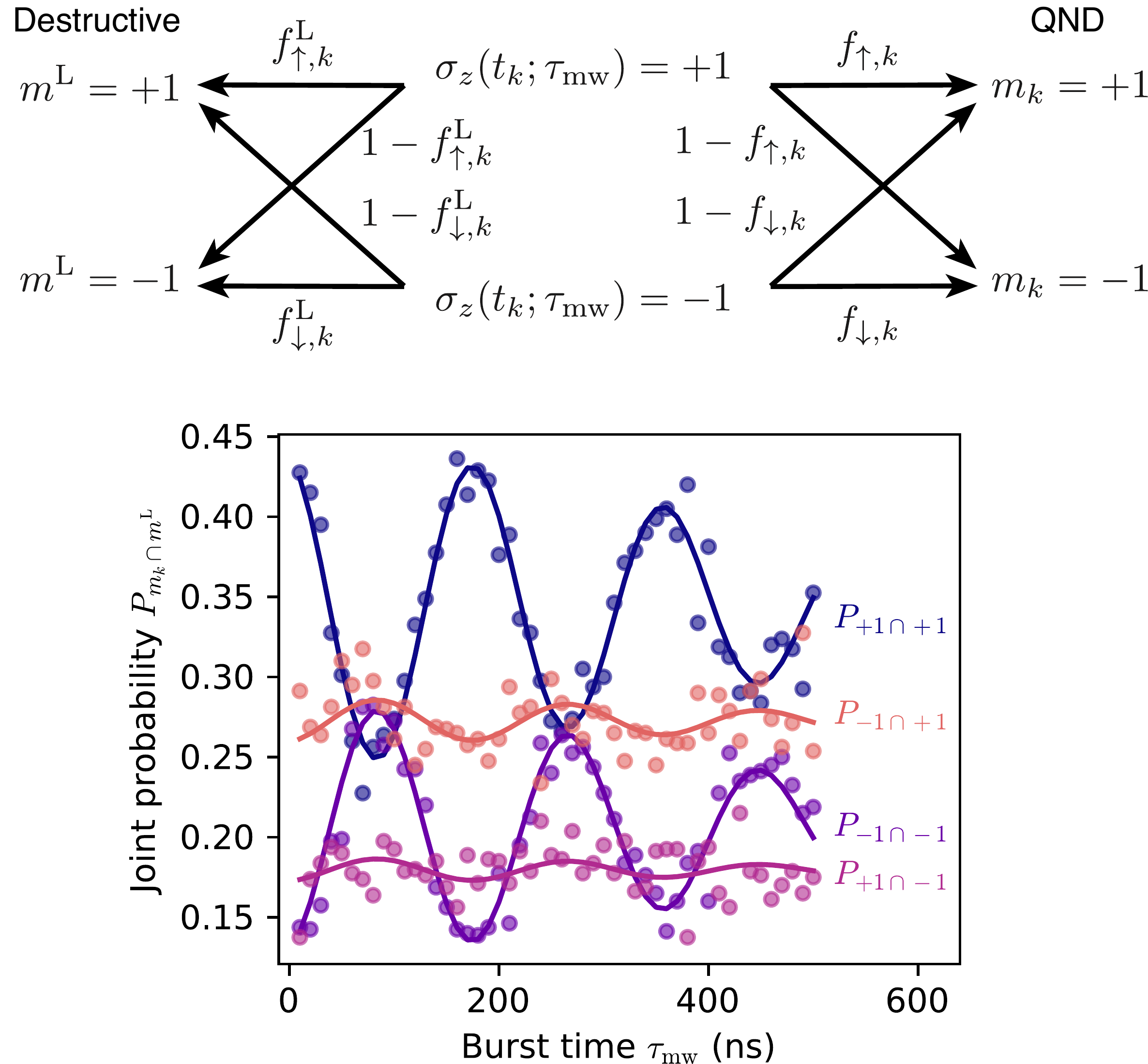}
\par\end{centering}

\protect\caption{\textbf{Correlation of measurement outcomes in two different schemes.}
The upper panel shows the probability flow for a given spin state $\sigma_z(t_k;\tau_\text{mw})$ in each single-shot measurement sequence.
When the up-spin (down-spin) state $\sigma_z(t_k;\tau_\text{mw})=+1$ ($-1$) is given at time $t_k$, it results in the QND readout estimator $m_k=+1$ ($-1$) with probability $f_{\uparrow,k}$ ($f_{\downarrow,k}$) and the destructive readout outcome $m^\text{L}=+1$ ($-1$) with probability $f^\text{L}_{\uparrow,k}$ ($f^\text{L}_{\downarrow,k}$).
The lower panel shows the probabilities of finding joint outcomes for $m_k$ and $m^\text{L}$, extracted from the same data set as the one used in Fig.~\ref{fig:fidelity}a ($k=5$). The offset of $P_{+1 \cap -1}(t_k)$ is, for example, given by $\{f_{\uparrow,k} (1 - f^\text{L}_{\uparrow,k}) - (1 - f^\text{L}_{\downarrow,k}) f^\text{L}_{\downarrow,k}\} B(t_k) + (1 - f_{\downarrow,k}) f^\text{L}_{\downarrow,k}$.
}
\label{fig:jointprobability}
\end{figure}

\end{document}